\documentclass[fleqn,12pt,twoside]{article}
\usepackage{espcrc1}

\usepackage{graphicx}
\usepackage[figuresright]{rotating} 

\usepackage{amsmath}
\usepackage{amsfonts}
\usepackage{amssymb}
\usepackage{epsfig}

\title{Statistical Mechanics of jamming and segregation in granular media}

\author{M. Nicodemi\address[napoli] 
{Dipartimento di Fisica, Universit\`{a} di Napoli``Federico II'',
INFM, Unit\`{a} di Napoli, Complesso Universitario Monte Sant'Angelo, Via
Cinthia, I-80126, Napoli, Italy},
A. Coniglio\addressmark[napoli],
A. de Candia\addressmark[napoli],
A. Fierro\addressmark[napoli], 
M. Pica Ciamarra\addressmark[napoli], 
and M. Tarzia\addressmark[napoli]}

\begin{document}
 
\maketitle

\begin{abstract} 
{\em In the framework of schematic hard spheres lattice models we discuss 
Edwards' Statistical Mechanics approach to granular media. 
As this approach appears to hold here to a very good approximation, 
by analytical calculations of Edwards' partition function at a mean field 
level we derive the system phase diagram and show that ``jamming'' 
corresponds to a phase transition from a ``fluid'' to a ``glassy'' phase, 
observed when crystallization is avoided. 
The nature of such a ``glassy'' phase turns out to be the
same found in mean field models for glass formers.
In the same context, we also briefly discuss mixing/segregation phenomena
of binary mixtures: the presence of fluid-crystal phase transitions 
drives segregation as a form of phase separation and, within a given phase, 
gravity can also induce a kind of ``vertical'' segregation, usually 
not associated to phase transitions.}
\end{abstract}

\section{Introduction}

We review some recent results in the study of glasses and dense granular 
materials obtained with the aid of schematic lattice models, 
simple enough to allow a full numerical and analytical investigation. 
Glasses and granular media 
exhibit deep similarities in their ``jamming'' behaviors, but although the 
idea of a unified description is emerging (see Silbert, O'Hern, Liu and 
Nagel contribution to this volume), the precise nature of jamming 
in non-thermal systems, such as granular media, and the origin of 
its close connections to glassy phenomena in thermal ones are still 
open and very important issues. 

An important conceptual open problem concerning granular media, is the 
absence of an established theoretical framework where they might be described. 
Edwards \cite{Edwards} (see Edwards, Bruji\'c, and Makse contribution to 
this volume) proposed a solution to such a problem by introducing 
the hypothesis that time averages of a system, exploring its mechanically 
stable states subject to some external drive (e.g., ``tapping''), coincide 
with suitable ensemble averages over its ``jammed states''.
We discuss here some recent results validating and generalizing 
Edwards' proposal 
(see also Dean and Lefevre, and Tarjus and Viot contributions). 
In particular, within this approach we show that mean field models 
for granular media undergo a phase transition from a (supercooled) 
``fluid'' phase to a ``glassy'' phase, when their crystallization 
transition is avoided.
The nature of such a ``glassy'' phase results to be the same found
in mean field models for glass formers: a discontinuous one step
Replica Symmetry Breaking phase preceded by a dynamical freezing
point. These results are supported by Monte Carlo (MC) ``tap dynamics'' 
simulations which, in the region of low MC shaking amplitudes, show a 
pronounced jamming similar to the one found in experiments on granular media
(see Bideau, Philippe, Ribi\`ere and Richard, and Caballero, Lindner, 
Ovarlez, Reydellet, Lanuza and Clement, and D'Anna and Mayor
contributions to the volume). 
As an application of Edwards' approach to powders we also briefly discuss 
segregation/mixing 
phenomena in these systems and discuss close correspondences 
with experiments (see Reis, Mullin and Ehrhardt contribution). 

Sec.\ref{granular} gives an short introduction to the phenomenology of 
granular materials relevant to the present purpose. 
Sec.\ref{sam} briefly reviews the essential 
lines in Edwards' Statistical Mechanics of powders and discuss some 
correspondence with other systems ``frozen'' at $T=0$, such as glasses 
in their inherent structures. 
In Sections \ref{mgm} to \ref{twotemp} we describe our results obtained 
in the study of hard spheres lattice models of granular materials, 
in order to check and apply Edwards' Statistical Mechanics to 
jamming and segregation. These are obtained by computer simulations 
and by analytic calculations with the generalized Bethe-Peierls method, 
or ``cavity method'', recently developed by M\'ezard and Parisi \cite{MP0102}. 
Finally, we draw some conclusions. 

\section{Granular Media}
\label{granular}

Granular media are large conglomerations of discrete macroscopic particles. 
Ubiquitous in the world around us, in many respect they behave
differently from any of the other familiar forms of matter 
- solids, liquids, or gases (for a review see \cite{JNBHM}).     
The static and dynamic properties of granular media, which are strongly 
dissipative systems, are not affected by temperature, because thermal 
fluctuations  are usually negligible. 
In sand the potential energy of a grain raised a distance equal to its own 
diameter can be $10^{12}$ times $k_B T$ at room temperature; therefore the 
temperature of the external bath can be considered zero and these media 
called {\em non-thermal}. As the system cannot explore its phase space, 
unless perturbed by external forces (such as shaking or tapping), 
it is frozen, at rest, in its mechanically stable microstates. 

Even though granular media may form crystalline packings, in most cases 
they are found at rest in disordered configurations, characterized by 
``fluid'' like distribution functions. Actually, gently shaken granular 
media exhibit a strong form of ``jamming'' 
\cite{Knight,Danna,Bideau}, i.e., an exceedingly slow dynamics, 
which shows deep connections to ``freezing'' phenomena observed 
in many thermal systems such as glass formers \cite{NCH,LN}. 
It was discovered by the Chicago group \cite{Knight} that under tapping 
the density of granular media tends to increase very slowly as a function 
of the number of shakes, in a stretched exponential \cite{Bideau} or even 
logarithmic \cite{Knight} way. 
These systems have typical ``aging phenomena'' along with 
huge relaxation times diverging \'a la Arrhenius or Vogel and Fulcher 
\cite{Knight,Danna,Bideau}, similarly to glass formers in the freezing region. 

Here, we discuss the nature of jamming in non-thermal systems \cite{LN,OHern}, 
and the origin of its close connections to glassy phenomena in thermal ones, 
in the framework of the Statistical Mechanics of powders introduced by 
Edwards \cite{Edwards,NCF1}: 
we derive the phase diagram of granular media and explain in a quantitative 
way their similarities and differences with glass formers. 
The introduction of a Statistical Mechanics for powders is grounded on 
observations from experiments \cite{Knight,Bideau} and 
simulations \cite{Makse,NCF1} that, as much as in thermal systems,
their macroscopic properties at stationarity are characterized by 
a few control parameters and their macrostates correspond to a huge 
number of microstates.

As an application of the Statistical Mechanics of powders, we also consider 
the intriguing phenomenon of segregation: in presence of shaking a granular 
system is not randomized, but its components tend to separate \cite{rev_segr}. 
An example is the so called ``Brazil nut'' effect (BNE) where, under shaking, 
large particles rise to the top and small particles move to the bottom of 
the container. Interestingly, by changing grains sizes or mass ratio or 
shaking amplitudes a 
crossover towards a ``reverse Brazil nut'' effect (RBNE) was more recently 
discovered \cite{luding} where small particles segregates to the top and large 
particles to the bottom. 
Several mechanisms have been proposed to explain these phenomena which, 
although of deep practical and conceptual relevance, are still largely 
unknown\cite{rev_segr}. 
Geometric effects, such as ``percolation'' \cite{rosato} or
``reorganization'' \cite{bridgewater,duran}, are known to be at work
since, in a nutshell, small grains appear to filter beneath large ones.
``Dynamical'' effects, such as convection \cite{knight} or
inertia \cite{shinbrot}, were shown to play a role as well.
Recent simulations and experiments have, however, outlined that 
segregation phenomena can involve ``global'' mechanisms, 
such as ``condensation'' \cite{luding} or, more generally, ``phase
separation'' \cite{kakalios}. We focus on these properties here. 


\section{Approaches \'a la Edwards to Statistical Mechanics 
of granular media} \label{sam}
In the Statistical Mechanics of powders introduced by Edwards \cite{Edwards} 
it is postulated that the system at rest (i.e., not in the ``fluidized'' 
regime) can be described by suitable ensemble averages over its 
``mechanically stable'' states (called here ``inherent states''). 
Edwards proposed a method to individuate the probability, $P_r$, 
to find the system in its inherent state $r$, under the assumption 
that these mechanically stable states have the same a priori probability 
to occur. 
The knowledge of $P_r$ has the conceptual advantage to substitute {\em time} 
with {\em ensemble averages}, allowing the description of the system 
properties by use of few basic theoretical concepts, as in thermodynamics. 
A possible approach to find $P_r$ is as follows \cite{Edwards,NCF1}. 
$P_r$ is obtained as the maximum of the entropy, 
\begin{equation}
S=-\sum_r P_r\ln P_r
\end{equation}
with the macroscopic constraint, in the case of the canonical ensemble, 
that the system energy, $E = \sum_r P_r E_r$, 
is given. This assumption leads to the Gibbs result:
\begin{eqnarray}
P_r\propto e^{-\beta_{conf} E_r}
\label{pr}
\end{eqnarray}
where $\beta_{conf}$ is a {\em Lagrange multiplier}, called 
{\em inverse configurational temperature}, enforcing the above constraint 
on the energy: 
\begin{equation}
\beta_{conf}= \frac{\partial S_{conf}} {\partial E}
\ \ \ \ \ \ \ \ \  S_{conf} = \ln \Omega_{IS} (E)
\end{equation}
Here, $\Omega_{IS}(E)$ is the number of inherent states with energy $E$. 
Thus, summarizing, the system at rest has 
$T_{bath} = 0$ 
and 
$T_{conf} =  \beta_{conf}^{-1} \neq 0$.

These basic considerations, to be validated by experiments or simulations, settle a theoretical Statistical Mechanics framework to describe granular media. Consider, for definiteness, a system of monodisperse hard spheres of mass $m$. In the system whole configuration space $\Omega_{Tot}$, we can write Edwards' generalized partition function as: 
\begin{eqnarray}
Z=\sum_{r\in\Omega_{Tot}} \exp(-{\cal H}_{HC}-\beta_{conf} mgH)
\cdot \Pi_r 
\label{Z_thedw}
\end{eqnarray}
where ${\cal H}_{HC}$ is the hard core interaction between grains, $mgH$ is the gravity contribution to the energy ($H$ is particles height), and the factor $\Pi_r$ is a projector on the space of ``mechanically stable'' states (i.e., inherent states space) $\Omega_{IS}$: if $r\in\Omega_{IS}$ then $\Pi_r=1$ else $\Pi_r=0$. 
Usual Statistical Mechanics, where $\beta_{conf}^{-1}$ is identified with $T_{bath}$, is recovered in the case where $\Pi_r = 1$ $\forall r$. 

In order to test Edwards proposal one introduces a dynamics (such as ``trapping'') that allows the system to explore the inherent states space. Using this dynamics one has first to check that at stationarity the system properties 
do not depend on the details of the dynamical history, i.e., a ``thermodynamic'' description is indeed possible. Then one must check that time averages obtained using such a dynamics compare well with ensemble averages over the distribution Eq.(\ref{pr}).

\subsection{Configurational entropy of granular media and glasses 
frozen at $T=0$} 
Edwards approach could also be relevant for a supercooled liquid
quenched at very low temperature (about zero). In this case the system remains blocked in its ``inherent structures'', i.e., the local minima of the potential energy in the particle configuration space 
\cite{Stillinger,parisi,sciortino,tartaglia}
(by analogy with the glass terminology we call ``inherent states'' the mechanically stable  states in granular materials). 

In glasses, in the inherent structure approach \cite{Stillinger,parisi,sciortino,tartaglia}, the configurational entropy (and consequently the configurational temperature) associated to the number of inherent structures corresponding to a given energy, $E$, can be defined too. In this context two different ways were essentially used to allow the system explore its inherent structure space. One way is by quenching  the system over and over from an equilibrium temperature, $T$, to zero temperature \cite{Stillinger,sciortino,tartaglia}.
Using this procedure, Sciortino {\em et al.}  \cite{sciortino} 
found that the configurational temperature numerically coincides with the equilibrium temperature $T$,  provided that $T$ is low enough.  
Another way to visit the inherent structures is by letting the system aging in contact with an almost zero bath temperature, $T_{bath}$. During the aging process an effective temperature, $T_{dyn}$, can be defined via the off-equilibrium extension of the Fluctuation-Dissipation ratio \cite{Kurchan}. It happens that in mean field models \cite{franzvirasoro,Kurchan} this effective temperature coincides with the configurational temperature, $T_{conf}$. This was also shown in Ref. \cite{barrat}, in the limit $T_{bath}\rightarrow 0$, for a class of finite dimensional systems.
The connection between Edwards approach for granular media and the results in glass theory has been pointed out in \cite{n1,barrat,Makse,NC,NCF1}. 

A second procedure, called ``tap dynamics'', similar to that used in the compaction of real granular materials was instead used in \cite{NC,NCF1,brey,Dean,berg,mehta,NCH}. In particular in Ref. \cite{NC,NCF1} each tap consists in raising the bath temperature to a value $T_{\Gamma}$ and, after a lapse of time $\tau_0$, quenching it back to zero. By cyclically repeating the process the system explores the space of the inherent states.  Once the stationary state is reached a temperature, $T_{fd}$, is defined via the equilibrium Fluctuation-Dissipation relation, and if Edwards' assumption applies, $T_{fd}$ coincides with the configurational temperature. This was verified for different finite dimensional models \cite{NC,NCF1}. Many other studies discussing Edwards' approach in this perspective were also presented \cite{brey,Dean,berg}.
In the following sections we give a picture of results \cite{NC,NCF1} 
about such a statistical mechanics approach in schematic models 
of granular media and glasses. 

\begin{figure}
 \begin{center}
 \includegraphics[angle=0, scale=0.7]{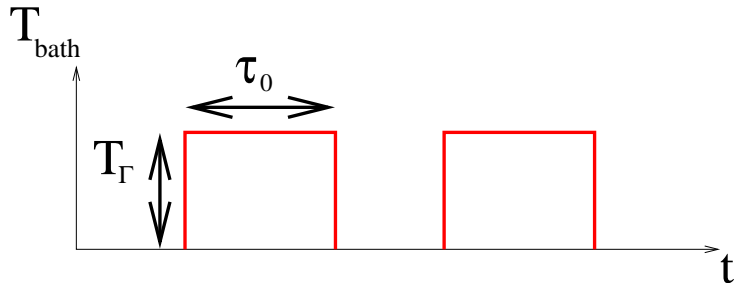}
 \caption{\label{tap} \footnotesize Our lattice models for granular media are subject to a Monte Carlo dynamics made of ``taps'' sequences. A ``tap'' is a period of time, of length $\tau_0$ (the tap duration), during which the system evolves at a finite bath temperature $T_\Gamma$ (the tap amplitude); after each ``tap'' the system evolves at $T_\Gamma = 0$ and reaches an inherent state.}
 \end{center}
\end{figure}

\section{Lattice models for granular media}
\label{mgm}

\subsection{A monodisperse hard-sphere system under gravity}
\label{hardsphere}
The simplest model for granular media we considered \cite{NCF1} is a system of hard-spheres of equal diameter $a_0=1$, subjected to gravity. We have studied this model on a lattice, constraining the centers of mass of the spheres on the sites of a cubic lattice (see inset in Fig.~\ref{universal}). 
The Hamiltonian of the system is:
\begin{equation}
{\cal H}= {\cal H}_{HC}(\{n_i\})+ gm \sum_i n_i z_i,
\label{Hhs}
\end{equation}
where the height of site $i$ is $z_i$, $g=1$ is gravity acceleration, $m=1$ the grains mass, $n_i=0,1$ the usual occupancy variable (i.e., $n_i=0$ or $1$ if site $i$ is empty of filled by a grain) and ${\cal H}_{HC}(\{n_i\})$ an hard-core interaction term that prevents the overlapping of nearest neighbor grains (this term can be written as ${\cal H}_{HC}(\{n_i\})=J\sum_{\langle ij\rangle }n_in_j$, where 
the limit $J\rightarrow\infty$ is taken). 

The grains are subject to a dynamics made of a sequence of Monte Carlo ``taps'' (see Fig. \ref{tap}): a single ``tap'' is a period of time, of length $\tau_0$ (the tap duration), where particles can diffuse laterally, upwards with probability $p_{up}\in[0,1/2]$, and downwards with probability $1-p_{up}$. 
When the ``tap'' is off grains can only move downwards (i.e., $p_{up}=0$) and the system evolves with $p_{up}=0$ until it reaches a blocked configuration 
(i.e., an ``inherent state'') where no grain can move downwards without violating the hard core repulsion. The parameter $p_{up}$ has an effect equivalent to keep the system in contact (for a time $\tau_0$) with a bath temperature $T_{\Gamma}=mga_0/\ln[(1-p_{up})/p_{up}]$ (called the ``tap amplitude''). 
The properties of the system are measured when this is in a blocked state. Time averages, therefore, are averages over the blocked configurations reached with this dynamics.
Time $t$ is measured as the number of taps applied to the system.

Under such a tap dynamics the systems reaches a stationary state where the Statistical Mechanics approach to granular media can be tested, and particularly Edwards hypothesis can be verified by comparing time averages to ensemble averages of Eq.(\ref{pr}).

\subsection{The stationary states of the tap dynamics}
\label{tapping}
During the tap dynamics, in the stationary state, the
time average of the energy, ${\overline E}$, and its fluctuations, ${\overline {\Delta E^2}}$, are calculated.
\begin{figure}
  \begin{tabular}{cc}
   \begin{minipage}[h]{0.45\textwidth}
 \begin{center}
\hspace{-2cm}\includegraphics[angle=-90,scale=0.35]{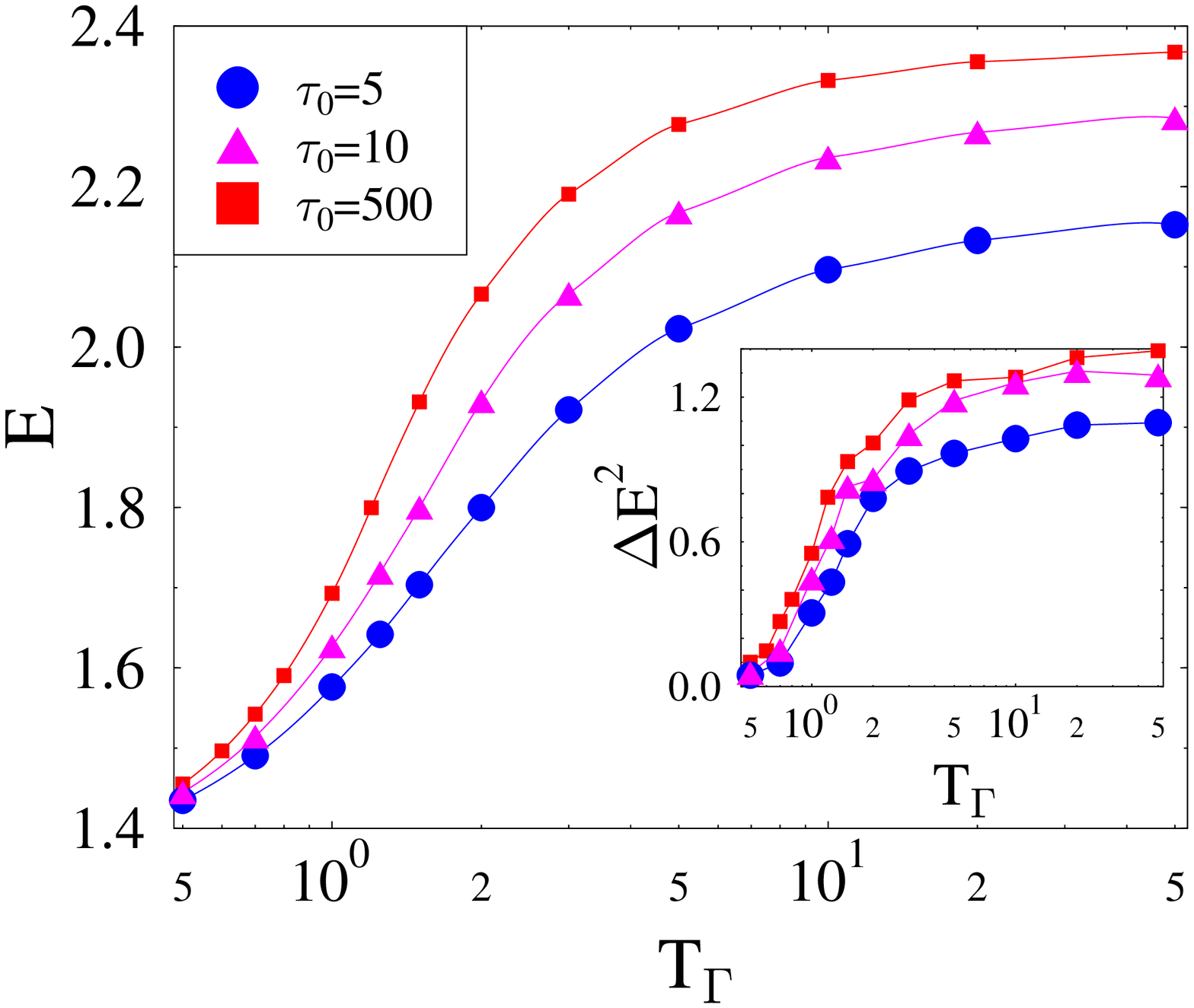}
\vspace{-2cm}
  \caption{\footnotesize
The time average of the energy, $\overline E$, and ({\bf inset}) its fluctuations, $\overline{\Delta E}^2$, recorded at stationarity during a tap dynamics, as a function of the tap amplitude, $T_\Gamma$, in the 3D lattice monodisperse hard sphere model.
Different curves correspond to sequences of tap with
different values of the duration of each single tap, $\tau_0$.
}
\label{energy}
 \end{center}
 \end{minipage}
&
   \hspace{0.5cm}

\begin{minipage}[h]{0.45\textwidth}
 \begin{center}
\hspace{-2cm}\includegraphics[angle=-90,scale=0.35]{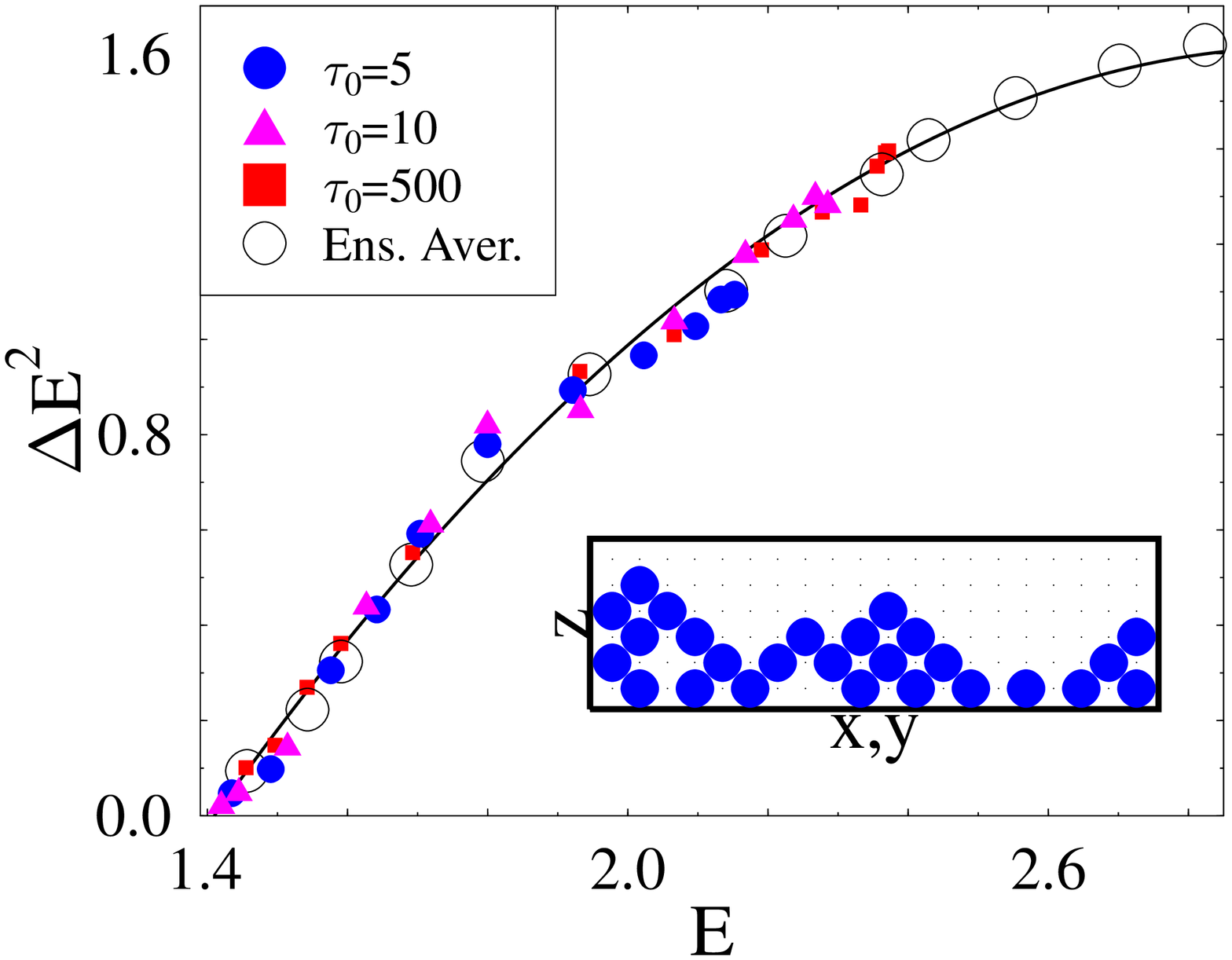}
\vspace{-2cm}
    \caption{\footnotesize
Time averages of energy fluctuations $\overline {\Delta E^2}$
plotted as function of the time average of energy $\overline {E}$.
$\bullet$, $\blacktriangle$ and $\blacksquare$ are time averages obtained with different tap dynamics. $\bigcirc$ are independently calculated ensemble averages according to Eq.(\ref{pr}).
The collapse of the data obtained with different dynamics shows that  the system stationary states are characterized by a {\em single} thermodynamic parameter. The agreement with the ensemble averages show the success of Edwards' approach to describe the system macroscopic properties.}
\label{universal}
 \end{center}
     \end{minipage}
 \end{tabular}
\end{figure}
Figure \ref{energy} shows $\overline{E}$ (main frame) and $\overline{\Delta E}^2$ (inset) as function the tap amplitude, $T_\Gamma$, (for several values of the tap duration, $\tau_0$). 
Since sequences of taps, with same $T_{\Gamma}$ and
different $\tau_0$, give different values of $\overline{E}$ and
$\overline{\Delta E}^2$, it is apparent that 
$T_{\Gamma}$ is not the right thermodynamic parameter. 
On the other hand, if the stationary states are indeed characterized by a {\em single} thermodynamic parameter 
the curves corresponding to different tap sequences (i.e. different
$T_{\Gamma}$ and $\tau_0$) should collapse onto a single master function, when $\overline{\Delta E}^2$ is parametrically plotted as function of $\overline E$. This is the case in the present model, where the data collapse is in fact found and shown in Fig. \ref{universal}. This is a prediction that could be easily checked in real granular materials.

A technique to derive from raw data the thermodynamic parameter $\beta_{fd}$ conjugated to $E$ (apart from an integration constant, $\beta_0$), is  through the usual {\em equilibrium} Fluctuation-Dissipation relation:
\begin{equation}
-\frac{\partial \overline E }{\partial \beta_{fd}}  =
\overline{\Delta E^2}.
\label{Sb}
\end{equation}
By integrating Eq.(\ref{Sb}), $\beta_{fd}-\beta_0$ can be expressed as function of $\overline E$ or (for a fixed value of $\tau_0$) as function of $\beta_\Gamma=1/T_\Gamma$: $\beta_{fd}-\beta_0\equiv g(\beta_\Gamma)$ (the constant $\beta_0$ can be determined as explained in \cite{NCF1}).
\begin{figure}
  \begin{tabular}{cc}
   \begin{minipage}[t]{0.45\textwidth}
    \begin{center}
    \hspace{-2cm}\includegraphics[angle=-90,scale=0.35]{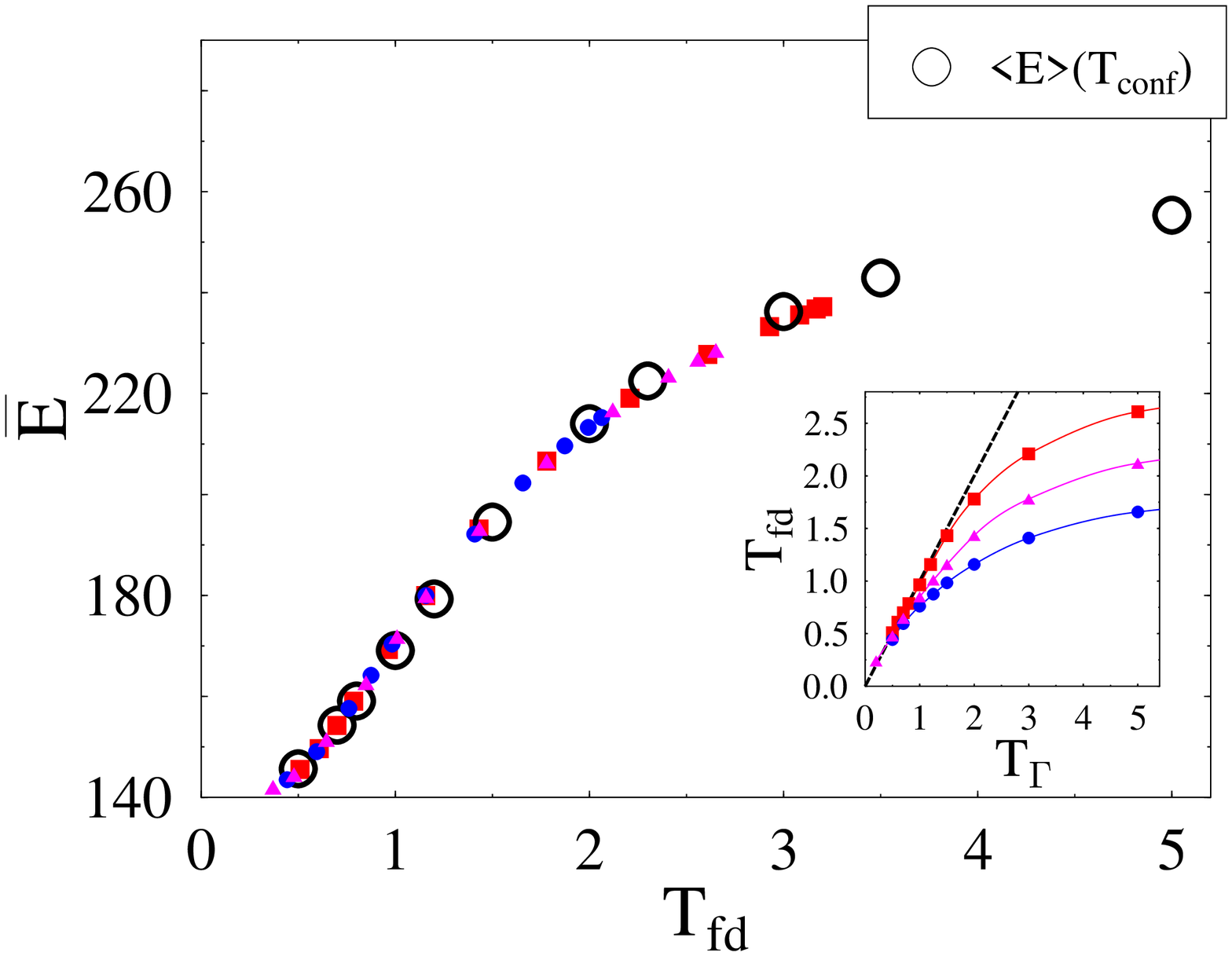}
\vspace{-2cm}
    \caption{\label{hard_sph}\footnotesize
{\bf Main frame}
The time average ${\overline E}$ and the ensemble average over the distribution Eq.(\ref{pr}) $\langle E \rangle$, plotted respectively as a function of $T_{fd}$  and $T_{conf}$ (in units $mga_0$), in the 3D monodisperse hard-sphere system under gravity described in the text. Symbols are as in Fig. \ref{universal}.
Time averages over the tap dynamics and Edwards' ensemble averages coincide. 
{\bf Lower Inset}
The temperature $T_{fd}\equiv \beta^{-1}_{fd}$ defined by
Eq.(\ref{Sb}) as function of $T_\Gamma$ (in units $mga_0$)
for $\tau_0=500,10,5~MCS$ (from top to bottom).
The straight line is the function $T_{fd}=T_\Gamma$.
}
\label{edwards}
\end{center}
   \end{minipage}
   \hspace{0.5cm}
&
\begin{minipage}[t]{0.45\textwidth}
\includegraphics[angle=-90,scale=0.35]{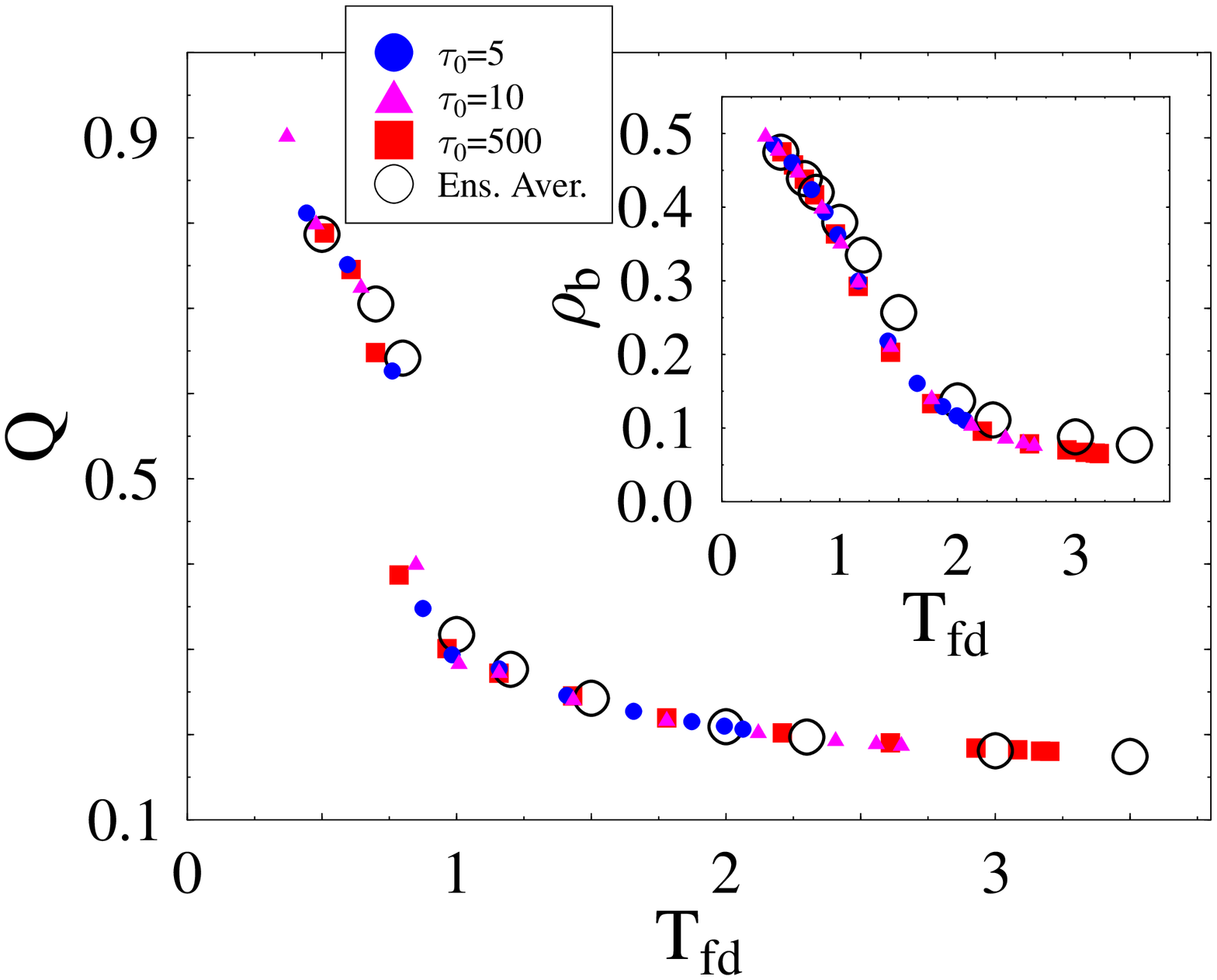}
\vspace{-2cm}
\caption{\label{mimmo}\footnotesize
The density self-overlap function, $Q$, and ({\bf upper inset})
the system density on the bottom layer, $\rho_b$,
plotted as function of $T_{fd}$ (in units $mga_0$), compared with the ensemble averages over the distribution Eq.(\ref{pr})
(the black empty circles), plotted as function of $T_{conf}$ (in units $mga_0$), in the 3D monodisperse hard-sphere system under gravity. 
Also for this system there is a very good agreement between the 
independently calculated time averages over the tap dynamics and the statistical mechanics ensemble averages \'a la Edwards. 
}
\end{minipage}
  \end{tabular}
\end{figure}

\subsection{Edwards' averages}
\label{EDW}
In Sect. \ref{tapping} we have found that the fluctuations of the energy in the stationary state depend only on the energy, $E$,
and not on the past history. More generally, we found \cite{NCF1} that all the 
macroscopic quantities we observed depend only on the energy, $E$, or on its conjugate thermodynamic parameter, $\beta_{fd}$, thus the stationary state can be  genuinely considered a ``thermodynamic state''. In this case one can attempt to construct an equilibrium statistical mechanics, as originally suggested by Edwards \cite{Edwards}.

To test Edward hypothesis one has to compare the time average of the energy, 
${\overline E}(\beta_{fd})$, recorded during the taps sequences,  with the ensemble average, $\langle E\rangle(\beta_{conf})$, over the distribution Eq.(\ref{pr}). 
Therefore we have independently calculated the ensemble average $\langle E\rangle$, as function of $\beta_{conf}$. 
Fig. \ref{edwards} (see also Fig. \ref{universal}) 
shows a very good agreement between $\langle E \rangle(\beta_{conf})$ and
${\overline E}(\beta_{fd})$ (notice that there are no adjustable
parameters).
In order to check the scenario in further details, we have also calculated the system density on the bottom layer, $\rho_b$,
and the density self-overlap function $Q$:
$Q = 1/N\sum_{i=1}^N n_i^{(1)}n_i^{(2)}$,
where the sum is over all the lattice sites and $n^{(1)}$ and $n^{(2)}$ refer to two different copies of the systems, which are at stationarity under the same tap dynamics.
We have verified that, when plotted as function of $T_{fd}$, both $\rho_b$ and $Q$ scale on a single master function (see Fig. \ref{mimmo}).
In Fig. \ref{hard_sph} (inset) we also show the dependence of the configurational temperature $T_{conf}$ on the parameters of the tap dynamics $T_\Gamma$ and $\tau_0$. 
Finally, we mention that we have also successfully tested Edwards scenario 
in an other model, the ``frustrated lattice gas'' \cite{dfnc,NCF1}, 
a system in the category of spin glasses. 

\subsection{Relaxation during MC ``tap'' dynamics}
The MC tap dynamics, in both the ``frustrated lattice gas'' and 
hard sphere models, exhibits a rich structure
in agreement with experimental findings \cite{Knight,Bideau}.
In the region of small tap amplitudes, the system gets ``jammed'' and
``memory'' phenomena are
observed, along with ``irreversibility'' effects \cite{Knight,NCH}.
In particular, it is interesting to consider correlation functions such as 
$C(t,t_w)=B(t,t_w)/B(t_w,t_w)$, where
$B(t,t_w)=\sum_i [\langle n_i(t+t_w)n_i(t_w)\rangle -\langle n_i(t+t_w)\rangle
\langle n_i(t_w)\rangle]$.
In the high $T_\Gamma$ region, $C(t,t_w)$ has a time translation invariant
(TTI) behavior, i.e., $C(t,t_w)=C(t)$. 
Asymptotically $C(t)$ can be well fitted by stretched exponentials:
$C(t)=C_0\exp[-(t/\tau)^\beta]$.
The exponent $\beta$ becomes significantly lower than 1 at low
amplitudes. The above fit defines the relaxation time
$\tau(T_\Gamma)$:
the growth of $\tau$ by decreasing $T_{\Gamma}$ is well approximated by
an Arrhenius or Vogel-Tamman-Fulcher law (as early found in \cite{NCH,NCF1}),
resembling the slowing down of glass formers close to the glass transition,
a result also recently experimentally reported in granular media
\cite{Danna,Bideau}:
$\tau\simeq \tau_0 \exp\left[E_0/(T_\Gamma-T_\Gamma^K)\right]$. 
The divergence point, $T_\Gamma^K$ (which in simulations is difficult to 
precisely locate; here it is between 
$0$ and $1$), of $\tau$ is interpreted as the
numerical location of the point of dynamical arrest of the system, where an ``ideal'' transition to a glassy phase occurs. By quenching the system at low values of $T_\Gamma$, the TTI character of relaxation is lost and logarithmic aging behaviors are found. For slow quenches the hard spheres model is able, anyway, to attain its crystal phase. The precise nature of the ``glassy'' region is, 
however, very difficult to be numerically determined, so in the next section 
an analytic approach is presented. 

An other dynamical quantity of interest is grains mean square displacement, 
$R^2(t)$, defining for $t\rightarrow\infty$ the diffusion
coefficient $D(\rho)=R^2(t)/t$. As much as in glass formers, 
$D$ goes to zero around a high density $\rho_c$, signaling a localization transition in which particles are confined in local cages and the macroscopic diffusion-like processes are suppressed \cite{NCH}. This phenomenon may also be described in a different way: $\rho_c$ is the density above which it becomes impossible to obtain a macroscopic rearrangement of particles positions without increasing the system volume, i.e., the density at which macroscopic shear in the system is impossible without dilatancy. This remark outlines the close correspondence with the phenomenon of the Reynolds transition in granular media \cite{NCH}.

\section{A mean field study of hard spheres under gravity on a random graph}
\label{mfgm}
\begin{figure}[ht]
\centerline{
\psfig{figure=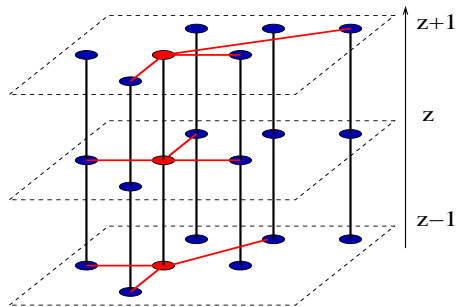,width=6cm,height=4cm,angle=0}
}
\caption{\footnotesize In our mean field approximation, the hard spheres describing
the system grains are located on a Bethe lattice, sketched in the figure,
where each horizontal layer is a random graph of given connectivity.
Homologous sites on neighboring layers are also linked and the overall
connectivity, $c$, of the vertices is $c\equiv k+1=5$.}
\label{Blattice}
\end{figure}

We have discussed up to here schematic models for granular media
shown to be well described by Edwards' assumption. 
Since the exact calculation of Edwards partition function, $Z$, 
for the above model of monodisperse hard 
spheres on a cubic lattice is hardly feasible, we now discuss a mean field 
theory \cite{cdfnt}. In an approximation \'a la Bethe-Peierls, we consider 
a random graph version of such a lattice, sketched in Fig. \ref{Blattice}: 
more specifically we consider a 3D lattice box with $H$ horizontal layers 
(i.e., $z\in\{1,...,H\}$) occupied by hard spheres; 
each layer is a random graph of given connectivity, $k-1$ 
(we take $k=4$); each site in layer $z$ is also 
connected to its homologous site in $z-1$ and $z+1$ 
(the total connectivity is thus $k+1$). 
The Hamiltonian is the one of Eq.(\ref{Hhs}) plus a chemical potential term to control the overall density. Hard Core repulsion prevents two connected sites to be occupied at the same time. 
In the present lattice model we adopt a simple definition of ``mechanical stability'': a grain is ``stable'' if it has a grain underneath. 
The operator $\Pi_r$ has thus a simple expression: 
$\Pi_r =\lim_{K\rightarrow\infty}\exp\left\{-K
{\cal H}_{Edw}\right\}$ 
where ${\cal H}_{Edw}=
\sum_i \delta_{n_i(z),1}\delta_{n_i(z-1),0}\delta_{n_i(z-2),0}$ 
(for clarity, we have shown the $z$ dependence in $n_i(z)$).

\begin{figure}[t!!!]
  \begin{tabular}{cc}
   \begin{minipage}[t]{0.45\textwidth}
    \begin{center}
\hspace{-1.8cm}   
    \includegraphics[scale=0.35,angle=-90]{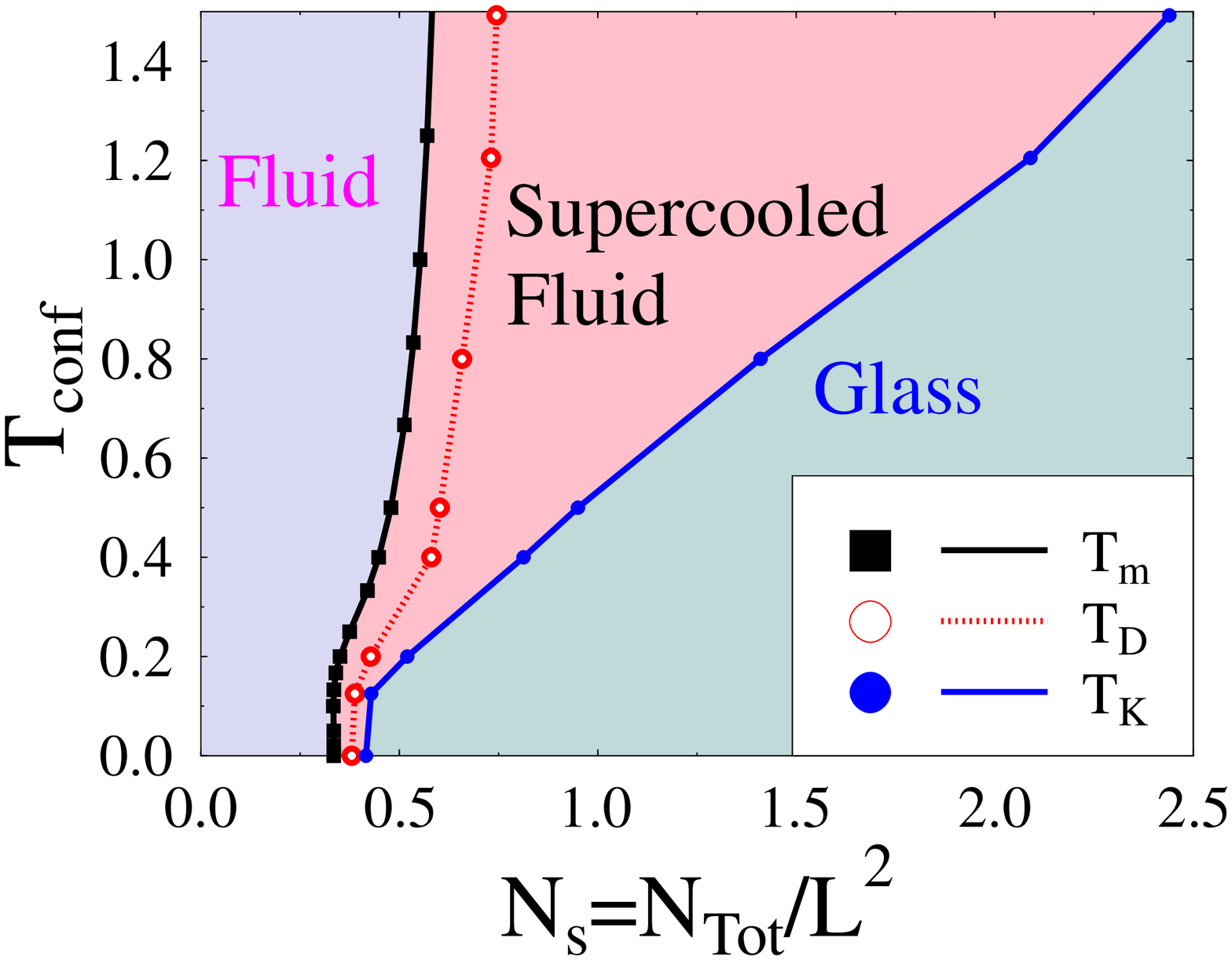}
\vspace{-2cm}
    \caption{\label{MFPD}\footnotesize
The system mean field phase diagram is plotted in the plane
of its two control parameters $(T_{conf},N_s)$: $T_{conf}$ is Edwards'
``configurational temperature'' and $N_s$ the average
number of grains per unit surface in the box.
At low $N_s$ or high $T_{conf}$, the system is found in a fluid phase.
The fluid forms a crystal below a melting transition line $T_m(N_s)$.
When crystallization is avoided, the ``supercooled'' (i.e., metastable)
fluid has a thermodynamic phase transition, at a point $T_K(N_s)$,
to a Replica Symmetry Breaking ``glassy'' phase with the same structure
found in mean field theory of glass formers. In between $T_m(N_s)$
and $T_K(N_s)$ a dynamical freezing point, $T_D(N_s)$, is located, where
the system characteristic time scales diverge.
}
\end{center}
   \end{minipage}
&
\begin{minipage}[t]{0.45\textwidth}
\hspace{-1cm}   
\includegraphics[angle=-90,scale=0.35]{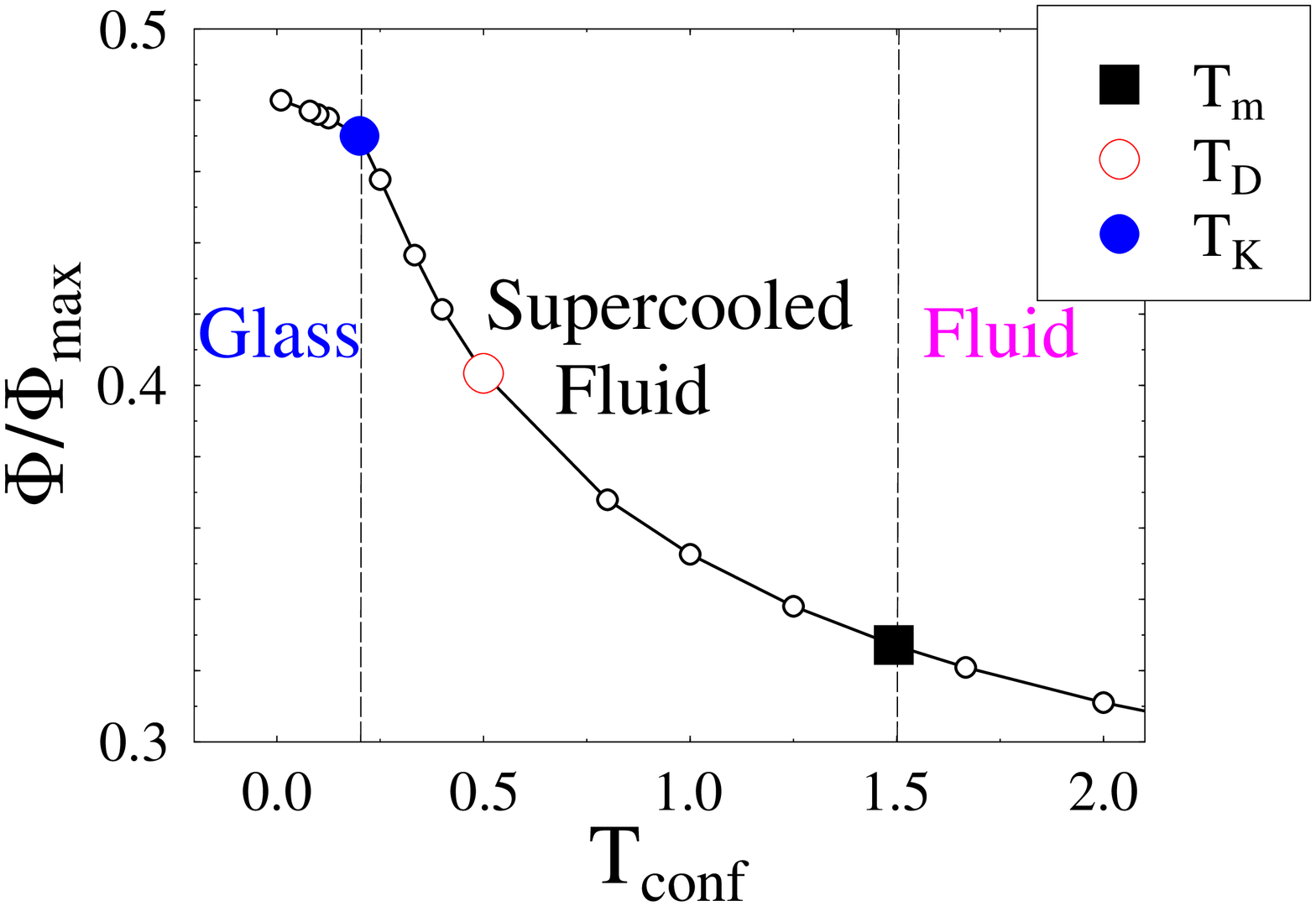}
\vspace{-2cm}
\caption{\label{phi_T}\footnotesize
For a system with a given number of grains (i.e., a given $N_s$),
the overall number density, $\Phi\equiv N_s/2\langle z\rangle$
($\langle z\rangle$ is the
average height), calculated in mean field approximation is plotted
as a function of $T_{conf}$; $\Phi(T_{conf})$ has a shape
very similar to the one observed in the ``reversible regime'' of
tap experiments and MC simulations of the cubic lattice model for
$\Phi(T_{\Gamma})$.
The location of the glass transition, $T_K$ (filled circle), corresponds to a cusp in the function $\Phi(T_{conf})$.
The passage from the fluid to supercooled fluid is $T_m$ (filled square).
The dynamical crossover point $T_D$ is found around
the flex of $\Phi(T_{conf})$ and well corresponds to the position
of a characteristic shaking amplitude $\Gamma^*$ found in experiments and simulations where the ``irreversible'' and ``reversible'' regimes approximately meet.
}
\end{minipage}
  \end{tabular}
\end{figure}

By using the ``cavity method'' \cite{MP0102}, the phase diagram is found
\cite{cdfnt}. 
At low $N_s$ or high $T_{conf}$ a fluid-like phase is found,
characterized by a homogeneous Replica Symmetric (RS) solution, in which only one pure state exists and the local fields are the same for all the sites of the lattice (translational invariance).
For a given $N_s$, by lowering $T_{conf}$ (see Figs. \ref{MFPD} and 
\ref{phi_T}), a phase transition to a crystal phase (an RS solution
with no space translation invariance) is found at $T_m$.
Notice that the fluid phase still exists below $T_m$ as a metastable phase corresponding to a supercooled fluid found when crystallization is avoided.
Within the one-step replica symmetry breaking  (1RSB) ansatz of the cavity method \cite{MP0102}, a non trivial solution  appears for the first time at a given temperature $T_D(N_s)$, signaling the existence of an exponentially high number of pure states.
In mean field theory $T_D$ is interpreted
as the location of a purely dynamical transition as in
mode-coupling theory, but in real
systems it might correspond just to a crossover in the dynamics
(see \cite{Kurchan,BM,Toninelli} and Ref.s therein).
The 1RSB solution becomes stable at a lower point $T_K$, where
a thermodynamic transition from the supercooled fluid
to a 1RSB glassy phase takes place (see Fig. \ref{MFPD}) in a scenario
\'a la Kauzmann with a vanishing complexity of pure states
(which stays finite for $T_K<T<T_D$).
 
The results of these calculations, summarized in the phase diagram of Fig. \ref{MFPD}, are further illustrated in Fig. \ref{phi_T}:
in a system with a given number of grains (i.e., a given $N_s$), the overall number density, $\Phi$, is plotted as a function of $T_{conf}$ (here by definition $\Phi\equiv N_s/2\langle z\rangle$, where $\langle z\rangle$ is the average height). The shown curve, $\Phi(T_{conf})$, is the equilibrium function here calculated. It has a shape very similar to the one observed in tap experiments \cite{Knight,Bideau}, or in MC simulations on the cubic lattice (see also \cite{NCH}), where the density is plotted as a
function of the shaking amplitude $\Gamma$ (along the so called 
``reversible branch''). In particular, a comparison of our mean field results 
with simulations of the 3D model 
of Hard Spheres under the tap dynamics shows a very good agreement. 

Summarizing, in the present mean field scenario of a granular medium with $N_s$ particles per surface, in general,
at high $T_{conf}$ (i.e. high shaking amplitudes) a fluid phase is located (see Fig. \ref{MFPD}).
By lowering $T_{conf}$, a phase transition to a crystal phase is found at $T_m$. However, when crystallization is avoided, the fluid phase still exists below $T_m$ as a metastable phase corresponding to a supercooled fluid.
At a lower point, $T_D$, an exponentially high number of new
metastable states appears, interpreted, at a mean field level,
as the location of a purely dynamical transition, which in real system is thought to correspond just to a dynamical crossover.
Finally, at a even lower point, $T_K$, the supercooled fluid has a
genuinely thermodynamics discontinuous phase transition to
glassy state. 
The structure of the glass transition of the present model for granular media, obtained in the framework of Edwards' theory,
is the same found in the glass transition of the $p$-spin glass and in other mean field models for glass formers \cite{Kurchan,BM}.

\section{A hard-sphere binary mixture under gravity}
\label{twotemp}
In order to study segregation and to test Edwards proposal to a more complicate system, here we consider a hard-sphere binary system made of two species 1 (small) and 2 (large) with grain diameters $a_0$ and $\sqrt{2} a_0$, under gravity
on a cubic lattice of spacing $a_0=1$.
We set the units such that the two kinds of grain have masses
$m_1=1$ and $m_2=2m_1$, and gravity acceleration is $g=1$. The hard core potential ${\cal H}_{HC}$ is such that two large nearest neighbor particles cannot overlap. This implies that only couples of small particles can be nearest neighbors on the lattice. The system overall Hamiltonian is:
\begin{equation}
{\cal H}={\cal H}_{HC}+ m_1gH_1 + m_2gH_2,
\label{HSM}
\end{equation}
where $H_1=\sum_{i}^{(1)}z_{i}$ and  $H_2=\sum_{i}^{(2)}z_{i}$, the
height of site $i$ is $z_i$ and the two sums are over all particles of species 1 and 2, respectively. In the above units, the gravitational energies in a given configuration are thus $E_1\;=\;H_1$ and $E_2\;=\;2H_2$.

\begin{figure}[t!!!]
  \begin{tabular}{cc}
   \begin{minipage}[t]{0.45\textwidth}
    \begin{center}
\hspace{-1cm}   
    \includegraphics[scale=0.35,angle=-90]{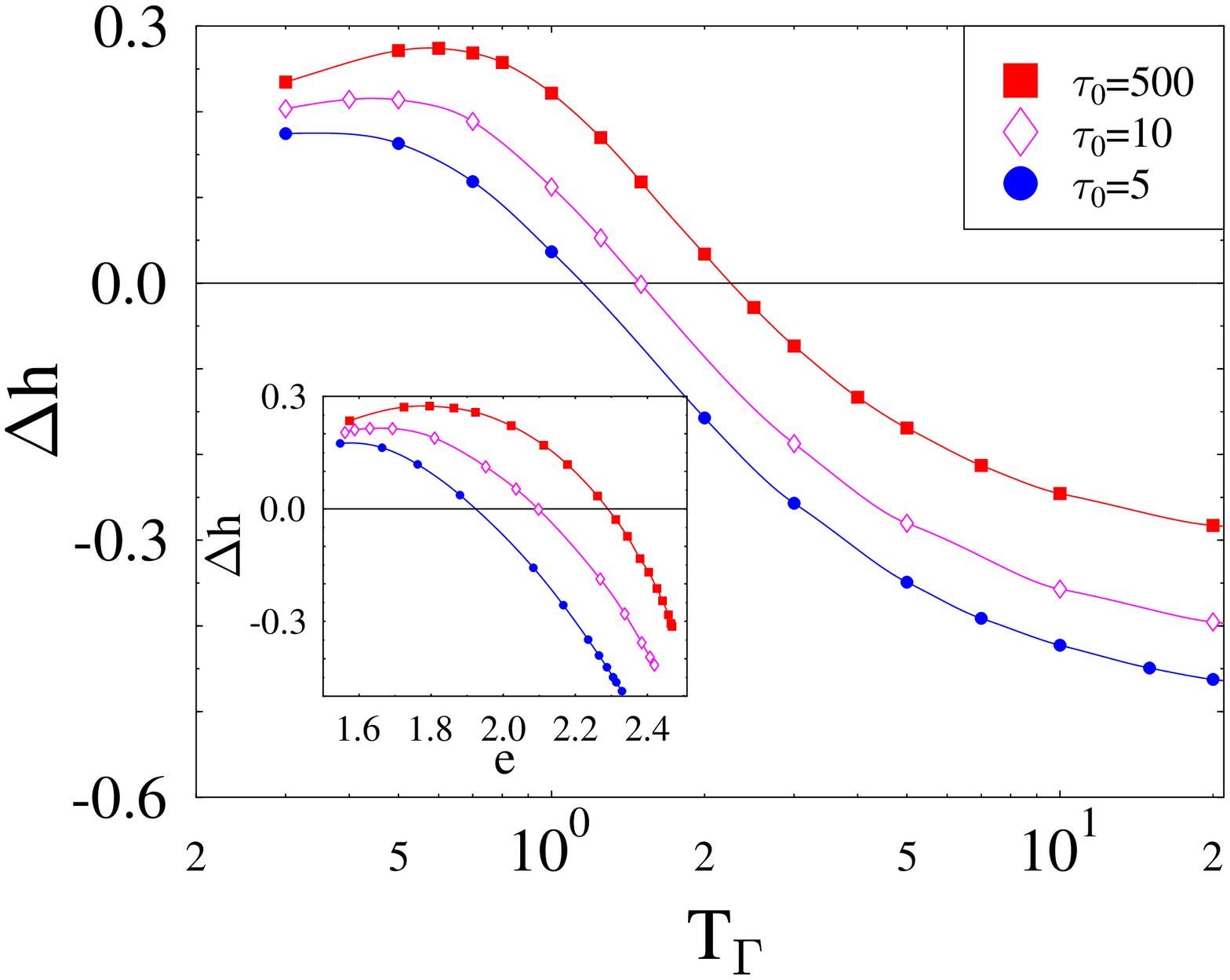}
\vspace{-0.5cm}
    \caption{\label{fig_dh}\footnotesize
{\bf Main frame} The difference of the average heights of small and large grains, $\Delta h=h_1-h_2$, measured at stationarity in the binary hard spheres mixture under gravity, 
is plotted as a function of tap amplitude, $T_{\Gamma}$
(in units $mga_0$). The three sets of points correspond to the shown tap durations, $\protect\tau_0$.
At high $T_{\Gamma}$ larger grains are found above the smaller, i.e, $\Delta h<0$, as in the Brazil nut effect (BNE).
Below a  $T^*_{\Gamma}(\tau_0)$ the opposite is found
(Reverse Brazil nut effect, RBNE).
{\bf Inset}
The $\Delta h$ data of the main frame are plotted as a function of the corresponding average energy, $e$. The three sets of data do not collapse, as before, onto a single master function.
}
\end{center}
   \end{minipage}
&
\begin{minipage}[t]{0.45\textwidth}
\hspace{0cm}   
\includegraphics[angle=-90,scale=0.35]{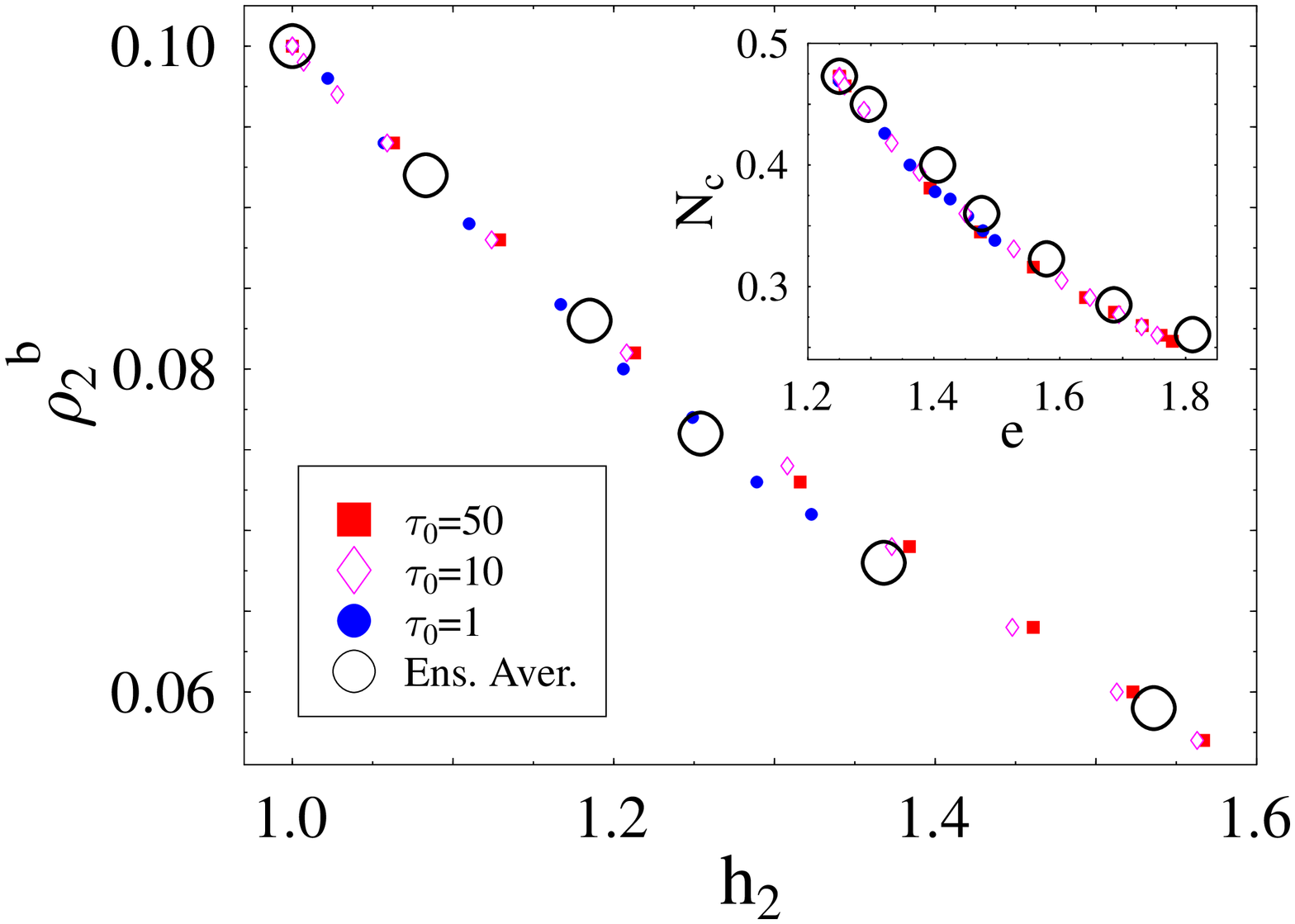}
\vspace{-0.5cm}
\caption{\label{fig_qr2nc_2}\footnotesize
{\bf Main frame} The average density of large grains on the box bottom layer, 
$\protect\rho_2^{b}$, measured at stationarity for
different $T_{\Gamma}$ and $\protect\tau_0$, scale almost on a single master
function when plotted as a function of the large grains height, $h_2$.
{\bf Upper inset} The average number of contacts between large grains
per particle, $N_c$, obtained for different $T_{\Gamma}$ and $\protect\tau_0$,
scale on a single master function when plotted as a function of the system
energy, $e$. 
}
\end{minipage}
  \end{tabular}
\end{figure}

As before, grains are confined in a box of linear size $L$ between with periodic boundary conditions in the horizontal directions and initially prepared in a random loose stable pack. Under the tap dynamics the system approaches a stationary state for each value of the tap parameters $T_{\Gamma}$ and $\tau_0$ used.
In Fig. \ref{fig_dh}, we plot as function of $ T_{\Gamma}$ (for several values
of $\tau_0$) the asymptotic value of the
{\em vertical} segregation parameter, i.e.,
the difference of the average heights of the small and large grains at
stationarity, $\Delta h(T_{\Gamma},\tau_0)\equiv h_1-h_2$.
Here $h_1$ and $h_2$ are the average of $H_1/N_1$ and $H_2/N_2$
over the tap dynamics at stationarity.

The results given in the main panel of Fig.~\ref{fig_dh}
apparently show that $T_{\Gamma}$ is not a right thermodynamic parameter, since sequences of taps with different $\tau_0$ give different values for the system observables.
However, if the stationary states corresponding to different tap dynamics (i.e., different $T_{\Gamma}$ and $\tau_0$) are indeed characterized 
by a single thermodynamic parameter, as in the monodisperse case above, 
the curves of
Fig. \ref{fig_dh} should collapse onto a universal master function when $\Delta h(T_{\Gamma},\tau_0)$ is parametrically plotted as function of an other macroscopic observable such as the average energy, $e(T_{\Gamma},\tau_0)=(E_1 + E_2)/N$ ($N$ is the total number of particles).
This collapse of data is not observed here, as it is apparent in the inset of Fig. \ref{fig_dh}.
We found, instead, \cite{NCF1} that two macroscopic quantities can be sufficient to characterize uniquely the stationary state of the system. These two quantities are, for instance, the energy $e$ and the height difference $\Delta h$. Of course since $e = ah_1 + 2bh_2$ (where $a= N_1/N$ and $b=N_2/N$) and $\Delta h = h_1-h_2$, we can also choose $h_1$ and $h_2$ to characterize the stationary state.
Namely, we found that a generic macroscopic quantity $A$, averaged over the tap dynamics in the stationary state, is only dependent on $h_1$ and $h_2$, i.e., $A= A(h_1,h_2)$. We have checked that this is the case for several independent observables, such as
the number of contacts between large particles, $N_{c}$, the density of  small and large particles on the bottom layer, $\rho_{1}^{b}$ and $\rho_{2}^{b}$, and others.
In particular, as shown in Fig. 
\ref{fig_qr2nc_2}, we find with good approximation that: $N_{c}\simeq N_{c}(e)=N_{c}(ah_1+bh_2)$,
$\rho_{2}^{b}\simeq\rho_{2}^{b}(h_{2})$, $
\rho _{1}^{b}\simeq \rho _{1}^{b}(h_{1})$.
Therefore we need both $h_{1}$ and $h_{2}$ 
to characterize unambiguously the state of the system;
namely all the observables assume the same values in a stationary
state characterized by the same values of $ h_{1}$ and $h_{2}$,
independently on the previous history (i.e., in our case independently on the
particular tapping parameters $T_{\Gamma}$ and $\tau_0$).

These findings imply that an extension of Edwards' approach is required, where at least {\em two} thermodynamic parameters have to be included \cite{NCF1}. As before, this can be obtained by assuming that the microscopic distribution is given by the principle of maximum entropy with the constraint that the average gravitational energies of the two species $E_1 =\sum_r P_r E_{1r} $ and $E_2 =\sum_r P_r E_{2r} $ are independently fixed.
This gives {\em two} Lagrange multipliers:
\begin{eqnarray}
\beta_1 = \frac{\partial \ln \Omega_{IS} (E_1,E_2) }{\partial E_1}\quad
\beta_2 = \frac{\partial \ln \Omega_{IS} (E_1,E_2) }{\partial E_2} \nonumber
\end{eqnarray}
\noindent
where $\Omega_{IS}(E_1,E_2)$ is the number of inherent states with $E_1$,$E_2$.

The hypothesis that the ensemble distribution at stationarity is the above 
can be tested as we have already previously shown. We have to check that
the time average of any quantity,
$A(h_{1},h_{2})$, as recorded during the taps sequences in a stationary state characterized by given values  $h_{1}$ and $h_{2}$, coincides with the ensemble average, $\langle A\rangle (h_{1},h_{2})$, over the generalized version of distribution Eq.(\ref{pr}). To this aim, we have calculated the
ensemble averages $\langle N_{c}\rangle $,  $\langle \rho_{2}^{b}\rangle$,
$\langle \rho _{1}^{b}\rangle $ for different values of $\beta_1$ and
$\beta_2$; we have expressed parametrically  $\langle N_{c}\rangle $, $\langle
\rho _{2}^{b}\rangle $, $\langle \rho _{1}^{b}\rangle $,
as function of the average of $ h_{1}$ and $ h_{2}$,
and compared them with the corresponding quantities, $N_{c}$, $\rho_{1}^{b}$
and $\rho _{2}^{b}$, averaged over the tap dynamics.
The two sets of data are plotted in Fig. 
\ref{fig_qr2nc_2} showing a
good agreement (notice, there are no adjustable parameters).

\vspace{-1.5cm}\begin{figure}[ht]
\centerline{
\hspace{-2cm}\psfig{figure=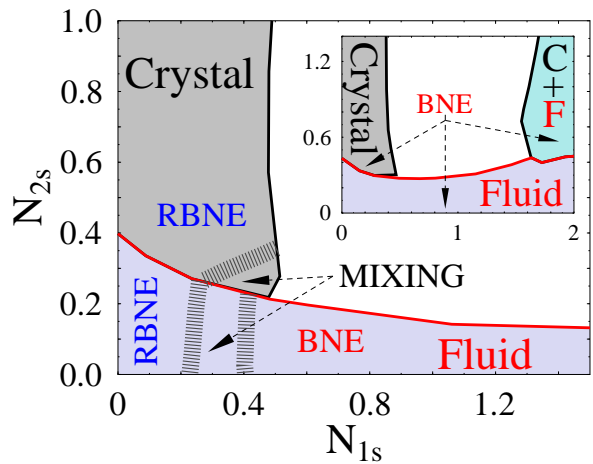,width=9cm,angle=-90}}
\vspace{-1.8cm}
\caption{\footnotesize 
Mean field phase diagram of the binary mixture in the plane 
$(N_{1s}, N_{2s})$ (particles densities per unit surface of species 1 and 2) 
for $m_1 \beta_1 = 0.8$ and $m_2 \beta_2 = 1.25$ 
({\bf main frame}). 
A fluid and a crystal phase are present (glassy and metastable 
phases are not shown here). Size segregation is found 
as a form of phase separation when the system is prepared out of these phases. 
Within a given phase, gravity drives vertical mixing/segregation phenomena 
(the BNE, RBNE and MIXING regions) separated by smooth crossovers 
(shaded areas). 
{\bf Inset} Phase diagram for $m_1 \beta_1 = 1.25$ and $m_2 \beta_2 = 0.8$. 
Here also a new phase, with strong BNE, appears where a fluid, 
rich in small grains, is found beneath a crystal rich in large ones 
(``C+F'' region). For $m_1 \beta_1 << m_2 \beta_2$, strong RBNE
can be found, with the fluid above the crystal. 
} \label{2pd}
\end{figure}

In \cite{CFNT} the mean field approximation of Sec. \ref{mfgm} is applied to the present binary mixture, giving a precise interpretation of segregation phenomena observed in the model of Eq.(\ref{HSM}), as shown in Fig.\ref{2pd}. 
In a nutshell, by changing $N_{1s}$, $N_{2s}$ (i.e., densities per unit 
surface of species 1 and 2), $\beta_1$ and $\beta_2$ (i.e, their configurational temperatures), or masses or sizes ratios, the system exhibits true
phase transitions from fluid to crystal phases. As much as in thermal media,
this induces segregation effects associated to phase separation phenomena,
with the formation of coexisting phases rich in small or large grains,
as experimentally observed in \cite{kakalios}. Within a given phase, 
the presence of gravity can also drive a form of ``vertical'' segregation,
where large grains are found on average above (the well known BNE) 
or below small grains (RBNE \cite{luding}), 
but this is usually not directly associated to phase transitions.

\section{Conclusions}
Summarizing, within the schematic framework of lattice models, 
we have shown, by Monte Carlo simulations of ``tap'' dynamics, 
that Edwards' approach to granular media at rest appears to be well grounded. 
The system stationary states are indeed independent on the sample history as in a ``thermodynamics'' system, and can be described in terms of a distribution function characterized by a few control parameters (such as configurational temperatures). 
By use of Edwards approach, we have derived, by analytical calculations
at a mean field level, the phase diagram of these systems. In particular, 
we discovered that ``jamming'' corresponds to a phase transition from 
a ``fluid'' to a ``glassy'' phase, observed when crystallization is avoided.
Interestingly, the nature of such a ``glassy'' phase turns out to be the
same found in mean field models for glass formers.
In the same framework, we have also briefly discussed segregation patterns 
observed in a hard sphere binary model under gravity subject to sequences 
of taps. Here, the presence of fluid-crystal phase transitions in the system 
drives segregation as a form of phase separation. Within a given phase, gravity
can also induce a kind of ``vertical'' segregation, not associated to phase
transitions.

In practice, even though the general validity of Edwards approach to ``frozen'' systems has just begun to be assessed, it turns out that a first reference framework is emerging to understand the physics of granular media and their deep connections with other ``jamming'' systems such as glass formers.

\end{document}